\documentclass[a4paper]{article}
 \usepackage{graphicx}     

\usepackage{amssymb}

\begin{document}
\tolerance=10000

\def\pni{\par \noindent}
\def\vsh{\smallskip}
\def\s{\smallskip}
\def\vs{\medskip}
\def\vvs{\bigskip}
\def\vvvs{\bigskip\medskip} 
\def\vsp{\vsh\pni}
\def\vsn{\vsh\pni}
\def\cen{\centerline}
\def\ra{\item{a)\ }} \def\rb{\item{b)\ }}   \def\rc{\item{c)\ }}
\def\eg{{\it e.g.}\ } 
\hyphenpenalty=2000

\font\title=cmbx12 scaled\magstep2
\font\bfs=cmbx12 scaled\magstep1
\begin{center}
 FRACALMO PRE-PRINT  \  {\tt www.fracalmo.org}
\\
 Applied Mathematics and Computation, 
\\ 
Vol. 187, No 1, pp. 295-305 (2007).
\vs

\hrule
\vvs

{\bfs{Some aspects of fractional diffusion  equations}}
\vs

{\bfs{of single and distributed  order}}
\footnote{
This paper is based on an invited talk given
by Francesco Mainardi
at the  International Symposium on {\it Analytic Function 
Theory, Fractional Calculus and Their Applications}, 
which took place at the University of
Victoria (British Columbia, Canada) on 22-27 August, 2005. The Symposium was 
organized in honor of Hari M. Srivastava, who celebrated his 65-th birthday in 2005.
Selected papers presented at the Symposium are published in this  special issue 
of AMC with  Guest Editors:
Hari M. Srivastava  and the Chairmen of the Organizing Committee
 Shigeyoshi Owa (Kinki University, Osaka, Japan),   Tadayuki Sekine (Nihon University, Chiba, Japan))
}
\vvs

Francesco MAINARDI$^{(1)}$, Gianni PAGNINI$^{(2)}$ and Rudolf GORENFLO$^{(3)}$
\vs

$^{(1)}$ Department of Physics, University of Bologna,
 and INFN, \\
   Via Irnerio 46, I-40126 Bologna, Italy\\
   {\tt francesco.mainardi@unibo.it}
\vs  

$^{(2)}$
National Agency for  New Technologies,
 Energy and the Environment,\\ 
 ENEA, Centre  "E. Clementel", 
 \\ Via Martiri di Monte Sole 4,
 I-40129 Bologna, Italy\\
 {\tt gianni.pagnini@bologna.enea.it}
 \vs
 
 $^{(3)}$Department of Mathematics  and Informatics,
 Free University of Berlin,\\
  Arnimallee  3, D-14195 Berlin, Germany \\
 {\tt gorenflo@mi.fu-berlin.de}
\end{center}

\tolerance=10000




\hyphenpenalty=2000



 \centerline{Dedicated to Professor H. M. Srivastava}
 \centerline{on the Occasion of his Sixty-Fifth Birthday}

\begin{abstract}The  time fractional diffusion equation
is obtained from the standard diffusion equation
by replacing the
 first-order time derivative with a fractional derivative
of order $\beta \in (0,1)\,.$
The fundamental solution for the {Cauchy} problem
is interpreted as
a probability density
of  a {self-similar} non-Markovian stochastic process
related to a phenomenon of sub-diffusion
(the variance grows in time  sub-linearly).
A further generalization is obtained by considering
a continuous or discrete  distribution of fractional
time derivatives of order  less than one.
 Then the fundamental solution is still a probability density
of a non-Markovian  process that, however, is no longer
 self-similar  but exhibits a corresponding distribution of time-scales.
\end{abstract}
\vs




\def\sg{\hbox{sign}\,}
\def\sgn{\hbox{sign}\,}
\def\sign{\hbox{sign}\,}
\def\e{{\rm e}}
\def\exp{{\rm exp}}
\def\ds{\displaystyle}
\def\dis{\displaystyle}
 \def\q{\quad}    \def\qq{\qquad} 
\def\lan{\langle}\def\ran{\rangle}
\def\lt{\left} \def\rt{\right}  
\def\lra{\Longleftrightarrow}
\def\d{\partial}
\def\dr{\partial r}  \def\dt{\partial t}
\def\dx{\partial x}   \def\dy{\partial y}  \def\dz{\partial z}
\def\rec#1{{1\over{#1}}}
\def\zr{z^{-1}}



\def\hatt{\widehat}
\def\epsilons{{\widetilde \epsilon(s)}}
\def\sigmas{{\widetilde \sigma (s)}}
\def\fs{{\widetilde f(s)}}
\def\Js{{\widetilde J(s)}}
\def\Gs{{\widetilde G(s)}}
\def\Fs{{\wiidetilde F(s)}}
 \def\Ls{{\widetilde L(s)}}
\def\L{{\mathcal L}} 
\def\F{{\mathcal F}} 
\def\M{{\mathcal M}} 
\def\P{{\mathcal{P}}} 
\def\H{{\mathcal{H}}} 

\def\NN{{\bf N}}
\def\RR{{\bf R}}
\def\CC{{\bf C}}
\def\ZZ{{\bf Z}} 


\def\I{{\cal I}}  
\def\D{{\cal D}}  


\def\erf{\hbox{erf}}  \def\erfc{\hbox{erfc}}


\def\uks{{\widehat{\widetilde {u}}} (\kappa,s)}

\def\psikappa{\psi_\alpha^\theta(\kappa)}



\section{Introduction}


The main physical purpose for adopting and investigating diffusion
equations of fractional order is  to
describe phenomena of  {\it anomalous diffusion} usually met in
transport processes through
complex and/or disordered systems including fractal media. In this
respect, in recent years   interesting reviews, see \eg
\cite{Metzler-Klafter JPhysics04,
Saichev PhysA05,
Zaslavsky PhysRep02},
have  appeared, to which (and references therein)
we refer the interested reader.
All the related models of random walk
turn out to be beyond the classical Brownian motion,
which is known to provide the microscopic foundation of
the standard diffusion,
see \eg 
\cite{
Klafter-Sokolov PHYSICSWORLD05,
SokolovKlafter EINSTEIN05}.
 \vsp
The diffusion-like  equations containing fractional derivatives
in time and/or in space are usually adopted
to model  phenomena of anomalous transport in physics, so
a detailed study of their solutions is required.
\vsp
   Our attention in this paper will be focused on the
time fractional diffusion equations of
a single or  distributed order less than 1,
which are known to be   models
for sub-diffusive processes.
Specifically, we have worked out how to express
their fundamental solutions in terms
of series expansions
 obtained from their representations
with Mellin-Barnes integrals.
\vsp
In Section 2 we shall recall the main results  for the
fundamental solution of the
time fractional diffusion equation of a single order,
which are obtained  by applying  two different strategies
in inverting its  Fourier-Laplace transform.
Both techniques yield   the same power series
representations
of the  required solution: it  turns out to be self similar
(through a definite space-time scaling relationship),
and  expressed in terms of
a special function  of the Wright type.
Then, in Section 3,  we  shall apply  the  second strategy
for obtaining the fundamental solution
in the general  case of a distributed order.
We have provided a representation in terms of
a Laplace-type integral of a  Wright function, 
that  can be expanded in  a series containing
 powers of  the space  variable   and certain functions of time, responsible
for  the time-scale distribution.
Finally, in Section 4, the main conclusions are drawn.
For convenience and self-consistency
we  provide an Appendix
devoted to our notations  of fractional calculus.
\section{The time fractional diffusion equation of single order}

The standard  {\it  diffusion equation}
in  re-scaled non-dimensional variables has the form
$$ \frac{\d}{\dt}\, u(x,t)
   =   {\d^2  \over \dx^2 } \,u(x,t)\,, \q
 x \in \RR \,,\q t \in \RR_0^+\,,
    \eqno(2.1)$$
        with $u(x,t)$ as  the field variable.
We assume  the initial condition
$$ u(x,0^+)= u_0(x)\,, \eqno(2.2)$$
where $u_0(x)$ denotes a given ordinary or generalized function
defined on $\RR$, that is Fourier transformable
in ordinary or generalized sense, respectively.
We assume to work in a suitable  space of generalized functions
where it is possible to deal freely
with delta  functions,  integral  transforms of
Fourier, Laplace and  Mellin type, and fractional integrals and derivatives.
\vsp
It is well known that the
 fundamental solution
(or {\it Green function}) of Eq. (2.1)
 i.e. the solution
subjected to the initial condition
$ u(x, 0) = u_0(x) =\delta (x)$,
and to the  decay to zero conditions  for $|x| \to \infty$,
is   
the Gaussian  {\it probability density  function} ($pdf$)
$$ u (x,t)
 = \frac{1}{2\sqrt{\pi }}\,t^{-1/2}\, \e^{-\ds x^2/(4t)}\,,
\eqno(2.3)$$
that evolves in  time with second moment
growing linearly with time,
$$  \mu _{2}(t) := \int_{-\infty}^{+\infty} \!\!\! x^{2}\,
  u(x,t)  \,dx = 2t\,, \eqno(2.4)$$
consistently with a law of {\it normal diffusion}\footnote{
The {centred second moment}
 provides the  {variance}  
usually denoted by $\sigma^2(t)$.
It is  a  measure for the spatial spread of  $u(x,t)$ with time
of a random walking particle starting at the origin $x=0$,
pertinent to the solution of the
diffusion equation (2.1) with initial condition $u(x,0) =\delta (x)$.
The asymptotic behaviour of the variance as $t\to \infty$
is relevant to distinguish {\it normal diffusion} ($\sigma^2(t)/t \to c > 0$)
from anomalous processes of {\it sub-diffusion} ($\sigma^2(t)/t \to 0$)
and of {\it super-diffusion} ($\sigma^2(t)/t \to +\infty $).}.
We note the {\it scaling property}  of the Green function,
expressed by the equation
$$ u(x,t) = t^{-1/2} \,  U(x/t^{1/2})\,, \q \hbox{with}\q
   U(x) := u(x,1) \,. \eqno(2.5)$$
The function $U(x)$ depending on the single variable $x$
turns out to be  an even function 
$U(x) = U(|x|)$
and is called the {\it reduced Green function}.
The  variable  $X :=x/t^{1/2}$
is known as the similarity variable.
It is  known that the  Cauchy problem $\{(2.1)-(2.2)\}$
is equivalent to the integro-differential equation
$$     u(x,t) = u_0(x) +
  {\ds \int_0^t}
 \lt[{\ds{\d^2 \over \dx^2}}\,u(x,\tau )\rt] \, d\tau\,,
    \eqno(2.6)$$
where the initial condition is incorporated.
        Now, by using the tools of the fractional calculus
        we can generalize the above Cauchy problem in order
        to obtain the so-called
 {\it time fractional diffusion equation}
in the two distinct (but mathematically equivalent)
forms available in the literature,
where the initial condition  is   understood as (2.2).
        For the essentials  of fractional calculus
we refer the interested reader to the Appendix.
        \vsp
        If $\beta$ denotes a real number such that $0 <\beta < 1$
the two forms are as follows:
        $$ \frac{\d}{\dt} u(x,t)
   =  \,_tD^{1-\beta}  \,   {\d^2  \over \dx^2 } \,u(x,t)\,,
\q  x \in \RR \,,\; t \in \RR_0^+\,; \q u(x, 0^+) = u_0(x)\,,
    \eqno(2.7)$$
where $\, _tD^{1-\beta}= \,_tD^1\,_tJ^\beta $
denotes the {\it Riemann-Liouville} (R-L) time-fractional derivative of
order $1-\beta$, see (A.4) with $m=1$,
and
$$ \,_tD_*^\beta  \, u(x,t)
   =   {\d^2  \over \dx^2 } \,u(x,t)\,, 
\q  x \in \RR \,,\; t \in \RR_0^+\,; \q u(x, 0^+) = u_0(x)\,,
    \eqno(2.8)$$
where $\,_t D_*^\beta=\,_tJ^{1-\beta} \,_tD^1 $
denotes  the  time-fractional derivative of order $\beta$
intended in the {\it Caputo} sense, see (A.5) with $m=1$.
In analogy with  the standard diffusion equation
we can provide an integro-differential form
that incorporates the initial condition (2.2): for this purpose
we replace in  (2.6) the ordinary integral with the
Riemann-Liouville fractional  integral  of order $\beta$,
$\,_tJ^{\beta}$,
 namely
 $$     u(x,t) = u_0(x) +
  \,_tJ^{\beta}
 \lt[{\ds{\d^2 \over \dx^2}}\,u(x,t\rt].
    \eqno(2.9)$$
Then, the above  equations read explicitly:
 $$ \frac{\d}{\dt} u(x,t)
   =
     {\ds{1\over \Gamma(\beta )}}  {\d  \over \dt }    \lt\{
  {\ds \int_0^t}
 \lt[{\ds{\d^2 \over \dx^2}}u(x,\tau )\rt] \,
  {d\tau \over (t-\tau)^{1-\beta}}  \rt\},
\; u(x, 0^+) = u_0(x)\,,
    \eqno(2.7')$$
$$  {\ds \rec{\Gamma(1-\beta )}} \,
  {\ds \int_0^t}   \lt[{\ds{\d \over \d \tau}} \, u(x,\tau)\rt]\,
  {\ds{d\tau \over (t-\tau)^{\beta}}}
   =   {\d^2  \over \dx^2 } \,u(x,t)\,,
\; u(x, 0^+) = u_0(x)\,,
    \eqno(2.8')$$
 $$     u(x,t) = u_0(x) +
   {\ds{1\over \Gamma(\beta )}}\,
  {\ds \int_0^t}
 \lt[{\ds{\d^2 \over \dx^2}}\,u(x,\tau )\rt] \,
  {d\tau \over (t-\tau)^{1-\beta}}  \,.
    \eqno(2.9')$$
The two Cauchy problems (2.7), (2.8) and the integro-differential equation (2.9)
are equivalent
\footnote{
The integro-differential equation (2.9) was investigated via
Mellin transforms
by Schneider \& Wyss \cite{SchneiderWyss 89}
in their pioneering 1989 paper.
The time fractional diffusion equation in the form (2.8) with the
Caputo derivative has been preferred and investigated by several authors.
From the earlier contributors let us quote Caputo
himself \cite{Caputo 69},
 Mainardi, see \eg
\cite{Mainardi WASCOM93,Mainardi CSF96,Mainardi CISM97}
 and Gorenflo et al. \cite{GorMaiSri Plovdiv97,Gorenflo-Rutman SOFIA94}.
In particular, Mainardi   has expressed the fundamental solution in terms of a
special function (of Wright type) of which he has studied the analytical
properties and provided plots also for $1<\beta <2$,
see also \cite{GoLuMa 99,GoLuMa 00,MainardiPagnini AMC03}
 and references therein.
For the form (2.7) 
with the R-L derivative 
earlier contributors include
the group of Prof. Nonnenmacher, 
see \eg \cite{Metzler PhysA94}, 
and Saichev \& Zaslavsky \cite{Saichev-Zaslavsky 97}.
}:
for example, we  derive (2.7) from (2.9)
 simply differentiating both sides of (2.9),
 whereas we  derive (2.9) from (2.8)  by fractional integration
 of order $\beta$. In fact,
 in view of the semigroup property (A.2) of the 
 fractional integral,   we note that
$$_tJ^\beta\, _tD_*^\beta u(x,t) \! = \!
  \,_tJ^\beta\, _tJ^{1-\beta}\, _tD^1 u(x,t) \! = \!
  \,_tJ^1\, _tD^1 u(x,t) \!= \! u(x,t)-u_0(x).\eqno(2.10)$$
In the limit  $\beta =1$ we recover  the well-known diffusion
equation (2.1).
\vsp
Let us consider from now on the  Eq. (2.8) with $u_0(x) = \delta(x)$:
the  fundamental solution can be obtained
by applying in sequence
the Fourier and Laplace transforms to  it.
We  write, for generic functions $v(x)$ and $w(t)$,
these transforms as follows:
$$
\begin{array}{ll}
&  \F \left\{ v(x);\kappa \right\} = \widehat v(\kappa)
  := \int_{-\infty}^{+\infty} \e^{\,\ds i\kappa x}\,v(x)\, dx\,,
  \;\kappa \in \RR\,, \\
&  \L \left\{ w(t);s \right\} = \widetilde w(s)
  := \int_{0}^{+\infty} \e^{\,\ds -st}\,w(t)\, dt\,,
  \;s \in \CC\,.
\end{array}
\eqno(2.11)
$$
Then,
in the Fourier-Laplace domain our  Cauchy problem
[(2.8) with $u(x,0^+) = \delta (x)$],
after applying   formula (A.6) for the
Laplace transform   of the fractional derivative 
and observing $\widehat \delta (\kappa ) \equiv 1$, see e.g. \cite{Gelfand-Shilov 64},
appears in the form
 $  s^\beta  \, \widehat{\widetilde{u}}(\kappa ,s) - s^{\beta -1}
    = -\kappa^2\,
   \widehat{\widetilde{u}}(\kappa ,s) \,,
$
from which we obtain
$$  \widehat{\widetilde{u}}(\kappa ,s)
   =  \frac{ s^{\beta -1}}{s^\beta + \kappa ^2 }\,, \q 0<\beta \le 1\,, \q
   \q \Re (s) > 0\,,\q \kappa \in \RR\,. \eqno(2.12)$$
To determine the  Green function $u(x,t)$
in the space-time domain we can follow two
alternative  strategies related to the
order in carrying out the inversions in (2.12).
\\
(S1) : invert  the Fourier transform
getting $\widetilde{u} (x,s)$
   and   then invert the remaining  Laplace transform;
\\
(S2) : invert  the Laplace transform getting $\widehat{u} (\kappa ,t)$
and then invert the remaining  Fourier transform.
\vsp
{\it Strategy (S1):}
The strategy (S1) has been applied by Mainardi
\cite{Mainardi WASCOM93,Mainardi CSF96,Mainardi CISM97}
to obtain the  Green function  in the form
$$ u (x,t) = t^{-\beta /2}\,
       U \lt(|x|/t^{\beta /2}\rt)\,,
\q -\infty < x <+\infty\,,  \q t\ge 0\,,
\eqno(2.13)
$$
where the variable
 $ X:=  x /t^{\beta /2}$ acts as {\it similarity variable}
and the function $U(x) := u(x,1)$ denotes
the {\it reduced Green function}.
Restricting from now on  our attention to $x\ge 0$,
the solution turns out   as
$$
\begin{array}{ll}
U(x)= {\ds \frac{1}{2}}\, M_{\frac{\beta}{2}} (x)
  = &
   {\ds \frac{1}{2}}\,    {\ds \sum_{k=0}^{\infty}\,
  \frac{(-x)^k}{ k!\,\Gamma[-\beta  k/2 + (1-\beta /2)]}} = \\
& {\ds \rec{2\pi}\, \sum_{k=0}^{\infty}\,\frac{(-x)^k }{ k!}\,
  \Gamma[(\beta  (k+1)/2]  \,\sin [(\pi \beta (k+1)/2]}
\,,
\end{array}
\eqno(2.14)$$
where $M_{\frac{\beta}{2}} (x)$
is an an entire
 transcendental function (of order $1/(1-\beta /2)$)
of the Wright type,
see also \cite{GoLuMa 99,GoLuMa 00} and  \cite{Podlubny 99}.
\vsp
{\it Strategy (S2):}
The strategy (S2) has been followed
 by Gorenflo et al.  \cite{GoIsLu 00} and
by Mainardi et al. \cite{Mainardi LUMAPA01}
to obtain the  Green functions of the more general
space-time fractional diffusion equations,
and requires to invert the Fourier transform
by using the machinery of the Mellin  convolution
and the   Mellin-Barnes integrals. 
Restricting ourselves here to recall the final results,
 the reduced Green function
for the time fractional diffusion equation
 now appears, for $x \ge 0$,  in the form:
$$ U(x)
 = \rec{\pi}\int_{0}^{\infty}
\!\! \cos\,(\kappa  x)  \,
  E_{\beta} \lt(-\kappa^2\rt)\,
d\kappa =
  \frac{1}{2x}\, \frac{1}{2\pi i}\,
   \int_{\gamma-i\infty}^{\gamma+i\infty}
  \!\!\frac{\Gamma(1-s)}{\Gamma (1- \beta s/2)}
 \, x^{\,\ds s}\,  ds\,,
 \eqno (2.15) $$
with $0 <\gamma< 1$,
where $E_\beta$ denotes the Mittag-Leffler function, see \eg
\cite{Erdelyi HTF,MaiGor JCAM00}.
By    evaluating the  Mellin-Barnes integrals using the residue theorem,
we arrive at the same power series (2.19).
\vsp
Both strategies allow us to prove that the Green function is
non-negative and normalized, so it can be interpreted
as a spatial probability density evolving in time with the similarity law
(2.13).
Although the two strategies are equivalent for yielding
  the required result,
 the second one appears more general and so more suitable
for treating the more complex case of fractional diffusion
of distributed order, see the next Section.
\vsp
Of  particular interest is the evolution of the  second moment
as it can be derived from Eq. (2.12) noting  that
$$   \widetilde\mu _2 (s)=
   - \frac{\d^2}{\d \kappa^2}\,
   \widehat{\widetilde {u}}(\kappa=0,s) =
 \frac{2}{s^{\beta +1}}
 \,, \q
 \hbox{so} \q
 \mu _2 (t)  = 2\,  \frac{t^\beta}{\Gamma(\beta +1)} \,,
 \eqno(2.16)$$
When  $0<\beta <1$
the sub-linear growth in time
is consistent  with   an anomalous  process   of  {\it  sub-diffusion}.

\section{The time-fractional diffusion equation of distributed order}

The fractional diffusion equation (2.8) can be generalized by using
the notion of  fractional derivative of distributed order
in time\footnote{
We find an earlier idea of fractional derivative of distributed order
in time in the 1969 book by Caputo \cite{Caputo 69},
that was later developed by Caputo himself,
see  \cite{Caputo FERRARA95,Caputo FCAA01}
and by Bagley \& Torvik, see \cite{BagleyTorvik 00}.}.
We now consider the so-called
{\it  time-fractional diffusion equation of distributed order}
$$
  \int _0^1 b(\beta )\,\lt[\, _tD_*^\beta  \, u(x,t)\rt] \, d\beta
\,    =
\frac{\d^2  }{\d x^2 }   \,u(x,t)\,,
\q  b(\beta)\ge 0, \;  \int _0^1 b(\beta )\,d\beta =  1\,,
\eqno(3.1) $$
with $  x\in \RR, \; t\ge 0$,
subjected to the initial condition $ u(x,0^+) = \delta  (x).$
 Clearly, some special conditions of regularity and boundary behaviour
      will be required for the weight function
       $b(\beta)$, that we call  the order-density.
\vsp
Equations of type (3.1) have  recently been discussed
in \cite{ChechkinGorenfloSokolov PRE02,ChechkinGorenfloSokolovGonchar FCAA03,%
ChechkinKlafterSokolov EUROPHYSICS03,SokolovChechkinKlafter 04}
and in \cite{Naber 04}.
As usual,we have considered  the initial condition $u(x,0^+) =\delta (x)$
 in order  to keep the probability meaning.
Indeed, already in  the  paper  \cite{ChechkinGorenfloSokolov PRE02},
it was shown that the  Green function
is non-negative and normalized, so
allowing interpretation as a density of the
probability at time $t$ of a diffusing particle to
be in the point $x$.
The main interest of the authors in
\cite{ChechkinGorenfloSokolov PRE02,ChechkinGorenfloSokolovGonchar FCAA03,%
ChechkinKlafterSokolov EUROPHYSICS03,SokolovChechkinKlafter 04}
was devoted to  the second moment of the Green function
(the displacement variance)  
in order to show the subdiffusive character
of the related stochastic process by analyzing
some interesting  cases of the order-density  function $b(\beta )$.
For a  thorough   general study of
fractional pseudo-differential equations of distributed order
let us cite the  paper
by Umarov and Gorenflo \cite{Umarov-Gorenflo ZAA05}.
For a   relationship with the Continuous Random Walk models
we may refer to the  paper by Gorenflo and Mainardi
\cite{GorMai CARRY04}.
\vsp
In this paper, extending the approach by Naber \cite{Naber 04},
 we  provide a general representation  of  the
fundamental solution corresponding
to   a generic 
order-density  $b(\beta )$.
By applying in sequence the Fourier and Laplace transforms
to Eq. (3.1) in analogy with the single-order case, see Eq. (2.12),
we obtain
 $$
\lt[\int _0^1 b(\beta )\, s^{\beta}\, d\beta \rt]
 \, \widehat{\widetilde{u}}(\kappa ,s)
 - \int _0^1 b(\beta )\,s^{\beta -1}\, d\beta
    = -\kappa^2\,
   \widehat{\widetilde{u}}(\kappa ,s) \,,
$$
from which
$$  \widehat{\widetilde{u}}(\kappa ,s)
   =  \frac{ B(s)/s}{B(s) + \kappa ^2}\,,
   \q \Re (s) > 0\,,\q \kappa \in \RR\,, \eqno(3.2)$$
where
$$ B(s) =  \int _0^1 b(\beta )\, s^\beta \,d\beta  \,. \eqno(3.3)$$
\vsp
Before  trying to get the solution in the space-time domain
 it is worth to outline the expression of its  second moment
as it can be derived from Eq. (3.2) using (2.16).
 We have, for $\kappa $ near zero,
$$ \widehat{\widetilde{u}}(\kappa,s) =
\frac{1}{s} \, \lt(1 - \frac{\kappa^2}{B(s)} + \dots \rt),
\; \hbox{so} \;
\widetilde{\mu _2}(s) =
- \frac{\d^2}{\d \kappa^2} \widehat{\widetilde{u}}(\kappa=0,s)=
  \frac{2}{s\, B(s)}.\eqno(3.4)$$
Then, from (3.4)  we are allowed to  derive the asymptotic behaviours
of $\mu _2(t)$ for $t\to 0^+$ and $t\to +\infty$
from the asymptotic behaviours of $B(s)$ for $s \to \infty$
and $s \to 0$, respectively, in virtue of the Tauberian theorems.
The expected  sub-linear growth  with time is shown in the
following special cases
of $b(\beta )$ treated in
 \cite{ChechkinGorenfloSokolov PRE02,ChechkinGorenfloSokolovGonchar FCAA03}.
The first case is {\it slow diffusion} (power-law growth)
where
$$ b(\beta ) = b_1 \delta (\beta -\beta _1)+ b_2 \delta (\beta -\beta _2),
\; 0<\beta_1 <\beta _2 \le 1, \; b_1>0, \; b_2 >0,\;
b_1 +b_2 =1.$$
In fact
$$   \widetilde{\mu _2}(s) = \frac{2}{b_1\,s^{\beta_1+1} + b_2 \,s^{\beta_2+1}},
\q
\hbox{so} \q
\mu _2(t)  \sim \cases{
 {\ds \frac{2}{b_2 \Gamma(\beta _2+1)}t^{\beta _2}}, &  $t \to 0,$ \cr\cr
{\ds \frac{2}{b_1 \Gamma(\beta _1+1)}t^{\beta _1}}, &  $t \to \infty.$}
\eqno(3.5)$$
In \cite{ChechkinGorenfloSokolov PRE02},
see  Eq. (16),
the authors were able to provide the analytical expression of
$\mu _2(t)$ in terms of a 2-parameter Mittag-Leffler function.
\vsp
The second case is  {\it super-slow  diffusion} (logarithmic growth)
where
$$ b(\beta ) = 1, \q  0\le \beta \le 1\,.$$
In fact
$$   \widetilde{\mu _2}(s) =
2\frac{\log s}  {s(s-1)} ,\q
\hbox{so} \q
\mu _2(t)  \sim \cases{
 2t\, \log (1/t), &  $t \to 0,$ \cr\cr
2\, \log (t),  &  $t \to \infty.$}
\eqno(3.6)$$
In \cite{ChechkinGorenfloSokolov PRE02},
see  Eqs. (23)-(26),
the authors were able to provide the analytical expression of
$\mu _2(t)$ in terms of an exponential integral  function.
\vsp
Inverting the Laplace transform, in virtue of
a theorem by  Titchmarsh  we obtain the remaining Fourier transform as
$$   \widehat {u}(\kappa, t) = -\rec{\pi }\,
\int_0^\infty \e^{-rt} \,  \hbox{Im}
\lt\{\widehat{\widetilde {u}} \lt(r \e^{i\pi }\rt)
\rt\} \, dr\,, \eqno(3.7) $$
that requires 
$ - \hbox{Im} \lt\{
B(s)/[s(B (s)+ \kappa ^2]\rt\}$
along the ray $s=r\,\e^{i\pi }$ with $r>0$
(the branch cut of the functions   $s^\beta $ and $s^{\beta -1}$).
By writing
$$ B \lt(r\, \e^{\,\ds i \pi}\rt) =
\rho \,\cos (\pi \gamma)
+ i \rho \sin (\pi \gamma),
\;
\cases{
{\ds  \rho =\rho (r) =\left\vert B\lt(r \,\e^{i\pi }\rt) \right\vert}\,, \cr
{\ds \gamma = \gamma (r) =
\rec{\pi}\,\hbox{arg}  \left[B\lt (r \,\e^{i\pi }\rt)\right]},
} \eqno(3.8)$$
 we get   after simple calculations
$$  \widehat {u}(\kappa, t) =
\int_0^{\infty}\,
\frac{\e^{\, \ds -rt}}{r} \,K  (\kappa,r) \,dr \,, \eqno(3.9)$$
where
$$ K(\kappa,r )
=    \frac{1}{\pi }\,
\frac{ \kappa ^2\rho \,\sin (\pi \gamma)}{\kappa^4 +
2 \kappa ^2 \, \rho\,\cos (\pi \gamma) +\rho^2}
\,. \eqno(3.10)$$
Since $u(x,t)$ is symmetric in $x$,
  Fourier inversion yields
$$
u(x,t)
=\frac{1}{\pi}\int_{0}^{+\infty} \cos(\kappa x) \,
\left\{\int_0^{\infty}
\frac{\e^{\, \ds -rt}}{r} \, \,K(\kappa,r )\, dr\right\} \, d\kappa\,.
\eqno(3.11)$$
To calculate this Fourier integral we use 
the Mellin transform.
Let
$$
   {\M} \, \{ f(\xi  ); s\} = f^*(s)=
   \int_0^{+\infty} f(\xi)\,
 \xi ^{\, \ds s-1}\,  d\xi ,  \q  \gamma_1< \Re (s) <\gamma_2,
                          \eqno(3.12)
$$
be the Mellin transform of a sufficiently well-behaved function
$f(\xi)\,,$ and let
$$
{\M}^{-1}\, \{  f^*(s ); \xi  \} =f(\xi )=
{1\over 2\pi i}\int_{\gamma -i
\infty}^{\gamma +i\infty} f^*(s)\, \xi ^{\, \ds -s} \,ds\,, \eqno(3.13)
$$
 be the inverse Mellin transform,
where $\, \xi  >0\,,$ $\, \gamma = \Re (s) \,,$
$\, \gamma_1< \gamma <\gamma_2\,.$
Denoting by   $\stackrel{{\M}} {\leftrightarrow}$
the juxtaposition of a function $f(\xi )$ with its
Mellin transform $f^*(s)$, the Mellin convolution theorem
implies
$$ h(\xi  ) =  f(\xi ) \otimes g(\xi ) :=
  \int_0^\infty
 \rec{\eta}\, f(\eta )\,g(\xi /\eta  )\,{d\eta  }
\,\stackrel{{\M}}{\leftrightarrow}\,
 h^*(s) = f^*(s)\,g^*(s) \,.\eqno(3.14) $$
Then, following  \cite{Mainardi LUMAPA01}, pp. 160-161,
 we identify the  Fourier integral in (3.11)
  as a  Mellin convolution in $\kappa $ ,
that is $u(x,t) =  f(\kappa, t) \otimes g(\kappa ,x)$,
if we set
(see  (3.14) with
  $\xi = 1/x$, $\eta =\kappa $)
 $$f(\kappa, t) :=
 \int_0^{\infty}\, \frac{\e^{\, \ds -rt}}{r} \, K(\kappa ,r) dr
       \,\stackrel{{\M}}{\leftrightarrow} \, f^*(s ,t)\,,
\eqno(3.15)$$
$$ g(\kappa,x ):= \rec{\pi\, x\, \kappa } \cos \lt( \rec{\kappa}\rt)
       \,\stackrel{{\M}}{\leftrightarrow} \,
{\Gamma(1-s)\over \pi \, x}
  \sin \lt( {\pi s\over 2}\rt)
 :=g^*(s,x)
 \,, \eqno(3.16) $$
with $0< \Re (s) <1$.
The next step thus consists in  computing the Mellin transform
$f^*(s,t)$ of the function $f(\kappa,t)$ and then  inverting the product
$ f^*(s,t) \, g^*(s,x)$ using  (3.16) in the  Mellin inversion formula,
namely
$$
\begin{array}{ll}
 u(x,t)
= &
{\ds \frac{1}{\pi x}
\frac{1}{2 \pi i} \int_{\sigma-i\infty}^{\sigma+i\infty}
f^*(s,t) \, \Gamma(1-s)\sin(\pi s/2) \, x^s \,ds} \\
 &  =
{\ds \frac{1}{x}
\frac{1}{2 \pi i} \int_{\sigma-i\infty}^{\sigma+i\infty}
f^*(s,t) \,
\frac{\Gamma(1-s)}{\Gamma(s/2)\Gamma(1-s/2)} \,x^s \,ds}
\,.
\end{array}
   \eqno(3.17)
$$
The required Mellin transform $f^*(s;t)$  is
$$
f^*(s,t)=\int_0^{\infty} \frac{\e^{-rt}}{r}\,
\left\{   \frac{1}{\pi}\,
\int_0^{\infty}
\frac{\kappa ^2\, \rho \sin(\pi\gamma)}
{\kappa ^4+2 \rho \cos(\pi \gamma ) \kappa ^2 + \rho^2} \,
\kappa ^{\, \ds s-1}\, d\kappa
\right\} dr
\,.
\eqno(3.18)$$
By the variable change  $\kappa ^2 \to \rho \mu$
the term in braces becomes
$$
\frac{\rho^{s/2+1}}{2 \rho} \,\frac{1}{\pi}
\int_0^{\infty}
\frac{\sin (\pi \gamma)}{\mu^2 + 2 \mu \cos(\pi \gamma) +1}
\, \mu^{(s/2+1)-1}\, d\mu = \qq \qq \qq \qq \qq \qq
\eqno(3.19)$$
$$ \qq \qq \qq =   -
\frac{\rho^{s/2}}{2}
\left\{
\frac{\Gamma(s/2+1)\,\Gamma[1-(s/2+1)]}
{\Gamma(\gamma s/2)\,\Gamma(1- \gamma s/2)}
\right\}
\,,
$$
where we  use a formula from  the Handbook by   Marichev,
see \cite{Marichev 83} p. 156, Eq. 15 (1),
under the condition
$0< \Re (s/2+1) < 2$, $|\gamma| < 1$.
As a consequence of (3.18)-(3.19)
 we  get
$$ f^*(s,t)=-\int_0^{\infty} \frac{\e^{\,\ds -rt}}{r}\,
   \frac{\rho^{\, \ds s/2}}{2} \,
\left\{
\frac{\Gamma(s/2+1)\,\Gamma[1-(s/2+1)]}
{\Gamma(\gamma s/2)\,\Gamma(1- \gamma s/2)}
\right\} \, dr
\,.\eqno(3.20)
$$
Now, using Eqs (3.17) and (3.20)  we can finally write the solution   as
$$
u(x,t)=\frac{1}{2\pi x}
\int_0^{\infty} \frac{e^{-rt}}{r}
F(\rho^{1/2}x)\, dr\,,   \eqno(3.21)
$$
where
$F(\rho^{1/2}x)$ is expressed in terms of Mellin-Barnes integrals:
$$
\begin{array}{ll}
F(\rho^{1/2}x)
= &
{\ds \frac{1}{2 \pi i} \int_{\sigma-i\infty}^{\sigma+i\infty}
\frac{\pi \Gamma(1-s)}
{\Gamma(\gamma s/2)\Gamma(1- \gamma s/2)}
(\rho^{1/2}x)^s \, ds}  = \\
& {\ds \frac{1}{2 \pi i} \int_{\sigma-i\infty}^{\sigma+i\infty}
\Gamma(1-s) \sin(\pi \gamma s/2)
(\rho^{1/2}x)^s\, ds} \,,
\end{array}
        \eqno(3.22)
$$
 with $\rho =\rho (r)$, $\gamma=\gamma (r)$. We remind
that   $\rho (r)$ and  $\gamma (r)$
are related to the order-density $b(\beta )$
according to Eqs. (3.3) and (3.8).
By evaluating the  Mellin-Barnes integrals via the residue theorem
we arrive at the series
representations in powers of $(\rho^{1/2}x)$,
$$
\begin{array}{ll}
F(\rho^{1/2}x)
= &
 {\ds \pi \rho^{1/2} x \,      \sum_{k=0}^{\infty}
\frac{(-\rho^{1/2}x)^k}
{k! \Gamma(\gamma k/2 +\gamma/2)
\Gamma(-\gamma k/2 +1 -\gamma/2)}}=  \\
& {\ds  \rho^{1/2} x \, \sum_{k=0}^{\infty}
\frac{(-\rho^{1/2}x)^k}
{k!} \sin(\pi \gamma (k+1)/2)}
 \,.
\end{array}
\eqno(3.23)
$$
This function can be interpreted again in terms of
generalized Wright  or $H$-Fox functions,
as outlined in \cite{Mainardi TEMME06}.
From  Eqs. (3.21) and (3.23),
interchanging  integration and summation,
we get
the series representation
of the fundamental solution:
$$
u(x,t)= \frac{1}{2 \pi}
\sum_{k=0}^{\infty}
\frac{(-x)^k}{k!} \,\varphi_k(t)\,, \eqno(3.24) $$
where, with $\rho =\rho (r)$, $\gamma =\gamma (r)$,
$$
\varphi_k(t)=
\int_0^{\infty} \frac{e^{-rt}}{r}
\sin [\pi \gamma (k+1)/2]\,
\rho^{(k+1)/2} \,dr \,.   \eqno(3.25)
$$
In order to check the consistency of this
general analysis
we find it instructive  to derive from Eqs. (3.24)-(3.25)
the results of Section 2
concerning the fractional subdiffusion of a single order.
We  denote  this (fixed) order by $\nu $
in distinction from the variable   $\beta $
in the distributed order  case.
This means to consider in Eq. (3.1)
 the particular case
$$ b(\beta)=\delta(\beta-\nu)\,,
 \quad 0<\nu < 1 \,,\eqno(3.26)$$
so  that  $ B(s) =  s^\nu $
and  Eq.  (3.8) yields
 $$ \rho =\rho(r)=r^{\nu}\,,\quad
\gamma= \hbox{const} =\nu \,.
\eqno(3.27)$$
In the series representation of the fundamental solution (3.24)-(3.25)
the functions  $\varphi_k(t)$   now  turn out as
$$
\begin{array}{ll}
 \varphi_k(t) = &
   {\ds \sin[\pi \nu (k+1)/2]\,
\int_0^{\infty} \frac{\e^{\,\ds -rt}}{r}\,
 r^{\, \ds \nu (k+1)/2}\, dr} = \\
& {\ds \sin[\pi \nu (k+1)/2]\, \frac{\Gamma [\nu(k+1)/2]} {t^{\nu(k+1)/2}}}
 \,.
\end{array}
\eqno(3.28)
 $$
As a consequence,
the solution reads
$$
\begin{array}{ll}
 u(x,t)= &
{\ds \frac{1}{2} t^{-\nu/2} \cdot
\frac{1}{\pi }\, \sum_{k=0}^{\infty}
\frac{(-x/t^{\nu/2})^k}{k!}\, \Gamma [\nu(k+1)/2]\,
\sin[\pi \nu (k+1)/2]} =\\
 & {\ds \frac{1}{2} t^{-\nu/2} \,
M_{\frac{\nu}{2}} \lt( \frac{x}{t^{\nu/2}}\rt)}\,,
\end{array}
\eqno(3.29)
$$
in agreement with Eqs. (2.13)-(2.14).
Of course, only in this special  case it is
possible to single out  a common  time factor  ($t^{-\nu/2}$)
from all the functions $\varphi_k(t)$
and get a self-similar  solution.
In  general the set of  functions $\varphi_k(t)$
gives rise to a distribution of different time scales related
to the order density $b(\beta )$.

\section{Conclusions} 
After  outlining the basic theory of the Cauchy problem
for the spatially one-dimensional and symmetric time fractional
diffusion equation (with its main equivalent formulations),
we have paid special attention to transform methods for
finding its fundamental solution or (exploiting self-similarity)
the corresponding reduced Green function.
We have  stressed the importance of the transforms of Fourier, Laplace and
Mellin and of the functions of Mittag-Leffler and Wright type,
avoiding however the cumbersome $H$-Fox function notations.
\vsp
A natural first step for construction of the fundamental solution
consists in applying in either succession the transforms of Fourier in space
and Laplace in time to the Cauchy problem.
This yields in the Fourier-Laplace domain the solution in explicit form,
but for the space-time domain we must invert both transforms in sequence
for which there are two choices, both leading to the same power series in the
spatial variable with time-dependent coefficients.
The strategy, called by us the "second",
 of first doing Laplace inversion and then the Fourier
inversion yields the reduced  Green function as a Mellin-Barnes integral
form  which, by the calculus of residues, the power series is obtained.
This strategy can be adapted to the treatment of the more
general case of the distributed order time fractional diffusion equation.
Now the fundamental solution can be expressed as an integral over
a Mellin-Barnes integral containing two parameters having the form of
functionals of the order-density.
Finally, again for the fundamental solution a power series comes out whose
coefficients, however, are time-dependent functionals of the order-density.
But, if there is more than one time derivative-order  present,
self-similarity is lost.

\section*{Appendix. The two  fractional derivatives}
For a sufficiently well-behaved function $f(t)$
($ t\in \RR^+$) we may define the fractional derivative
of order $\beta  $ ($m-1 <\beta \le m\,,$ $\, m\in \NN$),
 see \eg  \cite{GorMai CISM97,Podlubny 99}
in two different senses,  that we refer here as to
{\it Riemann-Liouville} derivative
and {\it Caputo} derivative, respectively.
Both derivatives are related to the so-called Riemann-Liouville
fractional integral of order $\alpha >0$, see
\cite{SKM 93,Srivastava-Saxena AMC03},
  defined as
$$ _t J^\alpha  \, f(t) :=    \rec{\Gamma(\alpha )}\,
\int_0^t (t-\tau)^{\alpha-1} \, f(\tau )\, d\tau\,, \q \alpha >0\,.
\eqno(A.1)  $$
We note the convention $\,_tJ^0 = I$ (Identity)
and the semigroup property
$$ _tJ^\alpha \, _tJ^\beta = \,
   _tJ^\beta  \, _tJ^\alpha = \, _tJ^{\alpha +\beta} \,, \q
 \alpha\ge 0,\; \beta \ge 0\,. \eqno(A.2)$$
The fractional derivative of order $\beta >0$ in the {\it
Riemann-Liouville} sense  is defined as the operator
$\,_tD^\beta$ which is the
left inverse of
the Riemann-Liouville integral of order $\beta $
(in analogy with the ordinary derivative), that is
$$ _tD^\beta \, _tJ^\beta  = I\,, \q \beta >0\,. \eqno(A.3) $$
If $m$ denotes the positive integer
such that $m-1 <\beta  \le m\,,$  we recognize from Eqs. (A.2) and (A.3)
$_tD^\beta  \,f(t) :=  \, _tD^m\, _tJ^{m-\beta}  \,f(t)\,, $
hence
$$
 _tD^\beta  \,f(t) = 
 \, \cases{
  {\ds {d^m\over dt^m}}\lt[
  {\ds \rec{\Gamma(m-\beta )}\int_0^t
    {f(\tau)\,d\tau  \over (t-\tau )^{\beta  +1-m}} }\rt] ,
 &  $m-1 <\beta  < m,$\cr\cr
   {\ds {d^m \over dt^m} f(t)} \,,
& $ \beta =m.$\cr}
\eqno(A.4)$$
For completion we define $ _tD^0 = I\,. $
On the other hand, the fractional derivative of order $\beta >0$ in the
{\it Caputo} sense  is defined as the operator
$\,_tD_*^\beta$  such that
$    _tD_*^\beta \,f(t) :=  \, _tJ^{m-\beta } \, _tD^m \,f(t)\,,$
hence
$$
    _tD_*^\beta \,f(t) =  
 \, \cases{
    {\ds \rec{\Gamma(m-\beta )}}\,{\ds\int_0^t
 {\ds {f^{(m)}(\tau)\, d\tau \over (t-\tau )^{\beta  +1-m}}}} \,,
& $ m-1<\beta  <m,$\cr\cr
   {\ds {d^m \over dt^m} f(t)} \,, & $ \beta =m.$\cr}
\eqno(A.5) $$
We point out the major utility
of the Caputo fractional derivative
in treating initial-value problems for physical and engineering
applications where initial conditions are usually expressed in terms of
integer-order derivatives. This can be easily seen
using the Laplace transformation, according to which
$$ \L \lt\{ _tD_*^\beta \,f(t) ;s\rt\} =
      s^\beta \,  \widetilde f(s)
   -\sum_{k=0}^{m-1}    s^{\beta  -1-k}\, f^{(k)}(0^+) \,,
  \q m-1<\beta  \le m \,,\eqno(A.6) $$
where
$ \widetilde f(s) =
\L \lt\{ f(t);s\rt\}
 = {\ds \int_0^{\infty}} \e^{\ds \, -st}\, f(t)\, dt\,, \;
s \in \CC$ and  $ f^{(k)}(0^+) := {\ds \lim_{t\to 0^+}}\, f^{(k)}(t)$.
The corresponding rule for the Riemann-Liouville
derivative is more cumbersome:  for $m-1<\beta  \le m $ it reads
$$ \L \lt\{ _tD^\beta  \, f(t);s\rt\} =
      s^\beta \,  \widetilde f(s)
   -\sum_{k=0}^{m-1}\,
\lt[_tD^k\, _tJ^{(m-\beta )}\rt]\,f(0^+) \, s^{m -1-k}\,,
\eqno(A.7)$$
where, in analogy with (A.6),  the limit for $t \to 0^+$
is understood to be taken after the operations of fractional integration
and derivation.  As soon as all the limiting   values $f^{(k)}(0^+)$
are finite
and $m-1 <\beta< m$, 
the formula (A.7)  simplifies into
$$ \L \lt\{ _tD^\beta  \, f(t);s\rt\} =
      s^\beta \,  \widetilde f(s) \,.\eqno(A.8)$$
In the special case   $f^{(k)}(0^+)=0$  for $k=0,1,  m-1$,
we recover the identity between the two fractional derivatives.
The Laplace transform rule (A.6)
was practically the starting point of Caputo \cite{Caputo 67,Caputo 69}
in defining his generalized derivative in the late sixties.
For further reading 
on the theory and applications of fractional calculus
we recommend
 the recent treatise
by  Kilbas, Srivastava \& Trujillo \cite{Kilbas-et-al BOOK06}.


\begin{thebibliography}{99}

 \bibitem{BagleyTorvik 00}
 R.L. Bagley and P.J. Torvik,
On the existence of the order domain and the solution of distributed
 order equations:  I and II,
{\it  Int. J. Appl. Math.} {\bf 2} 
(2000) 865-882 and  965-987.



\bibitem{Caputo 67}
M. Caputo,
  {Linear models of dissipation whose $Q$ is almost frequency
  independent,  Part II}.
  {\it Geophys. J. Roy. Astronom. Soc.} {\bf 13}  (1967) 529--539.

 \bibitem{Caputo 69}
 M. Caputo,
{\it Elasticit\`a e Dissipazione},
  Zanichelli, Bologna, 1969 (in Italian).

\bibitem{Caputo FERRARA95}
  M. Caputo,
  Mean fractional-order derivatives differential equations and filters,
  {\it Ann. Univ. Ferrara, Sez. VII, Sci. Mat.} {\bf 41} (1995) 73-84.


\bibitem{Caputo FCAA01}
 M. Caputo,
 Distributed order differential equations modelling
  dielectric induction and diffusion.
{\it Fract. Calc.  Appl. Anal.} {\bf 4} 
 (2001) 421-442.

\bibitem{ChechkinGorenfloSokolov PRE02}   
A.V. Chechkin, R. Gorenflo and I.M. Sokolov,
Retarding subdiffusion and accelerating superdiffusion
governed by distributed-order fractional diffusion equations,
{\it Phys. Rev. E} {\bf  66} (2002) 046129/1-6.
\bibitem{ChechkinGorenfloSokolovGonchar FCAA03}  
A. V. Chechkin, R. Gorenflo, I. M. Sokolov, V. Yu. Gonchar,
Distributed order time fractional diffusion equation,
{\it Fract. Calc.  Appl. Anal.} {\bf 6} (2003) 259-279.
\bibitem{ChechkinKlafterSokolov EUROPHYSICS03}   
A.V. Chechkin, J. Klafter and I.M. Sokolov,
Fractional Fokker-Planck equation for ultraslow kinetics,
{\it Europhys. Lett.} {\bf  63} (2003) 326-332.


 \bibitem{Erdelyi HTF}
  A. Erd\'elyi, W. Magnus, F. Oberhettinger and F.G. Tricomi,
  {\it Higher Transcendental Functions},
 Bateman Project,  Vols. 1-3,
  McGraw-Hill, New York, 1953-1955.
 [Vol. 3,  Ch. 18: Miscellaneous Functions, pp. 206-227]



\bibitem{Gelfand-Shilov 64}
 I. M. Gel\`{}fand and G. E. Shilov,
 {\it Generalized Functions}, Volume I.
 Academic Press, New York and London, 1964.


 \bibitem{GoIsLu 00}
 {R. Gorenflo, A. Iskenderov and Yu. Luchko},
 Mapping between solutions of fractional diffusion-wave equations,
 {\it Fract. Calc.   Appl. Anal.\/}
 {\bf  3}   (2000) 75--86.

\bibitem{GoLuMa 99}
{R. Gorenflo, Yu. Luchko and  F. Mainardi},
Analytical properties and applications of the Wright function.
{\it Fract. Calc.  Appl. Anal.} {\bf 2} (1999) 383-414.
\bibitem{GoLuMa 00}
{R. Gorenflo, Yu. Luchko and  F. Mainardi},
Wright functions as scale-invariant solutions of the diffusion-wave
 equation,
{\it J. Comput. and Appl. Mathematics\/}
     {\bf 118}  (2000) 175-191.


 \bibitem{GorMai CISM97}
  R. Gorenflo and F. Mainardi,  Fractional calculus:
  integral and differential equations of fractional order,
  in: A. Carpinteri and F. Mai\-nardi (eds.),
  {\em Fractals and Fractional Calculus in Continuum Mechanics\/},
  Springer Verlag, Wien and New York, 1997,  pp. 223--276.
  [Reprinted in  {\tt http://www.fracalmo.org}]

 \bibitem{GorMai CARRY04}
  R  Gorenflo  and  F. Mainardi,
    Simply and multiply scaled diffusion limits for continuous time
     random walks,
    in: S. Benkadda, X. Leoncini and  G. Zaslavsky (eds.),
   {Proceedings of the  International Workshop on
  Chaotic Transport and Complexity  in Fluids and Plasmas}
     Carry Le Rouet (France) 20-25 June 2004,
    {\it IOP (Institute of Physics) Journal of Physics: Conference
    Series} {\bf 7},  2005, pp. 1-16.

\bibitem{GorMaiSri Plovdiv97}
  R. Gorenflo, F. Mainardi and H.M. Srivastava,
      Special functions in fractional relaxation-oscillation and
      fractional diffusion-wave phenomena, in D. Bainov (Editor),
      Proceedings VIII International Colloquium on Differential Equations,
      Plovdiv  1997,
      VSP, 
      Utrecht, 1998, pp. 195-202.
\bibitem{Gorenflo-Rutman SOFIA94}
 R. Gorenflo and R. Rutman,
 On ultraslow and intermediate processes,
 in P. Rusev, I. Dimovski and V. Kiryakova (eds.),
 {\it Transform Methods and Special Functions, Sofia 1994},
 Science Culture Technology, Singapore, 1995, pp. 171-183.

 \bibitem{Kilbas-et-al BOOK06}          
 A.A. Kilbas, H.M. Srivastava  and J.J. Trujillo,
 {\it Theory and Applications of Fractional Differential Equations},
 North-Holland, Amsterdam, 2006.

 \bibitem{Klafter-Sokolov PHYSICSWORLD05}
 J. Klafter and I.M. Sokolov,
 Anomalous diffusion spreads its wings,
 {\it Phys. World} {\bf  18} (2005) 29-32.

 \bibitem{Mainardi WASCOM93}
  F. Mainardi,
  On the initial value problem for the fractional diffusion-wave equation,
 in: S. Rionero and T. Ruggeri (eds.),
 {\it Waves and Stability in Continuous Media},
  World Scientific, Singapore,  1994, pp. 246-251.


\bibitem{Mainardi CSF96}
  F. Mainardi,
  Fractional relaxation-oscillation and fractional
  diffusion-wave phenomena,
  {\it Chaos Solitons Fract.\/} {\bf 7} (1996) 1461--1477

 \bibitem{Mainardi CISM97}
   F. Mainardi,  Fractional calculus:
  some basic problems in continuum and statistical mechanics,
  in: A. Carpinteri and F. Mainardi (eds.),
  {\em Fractals and Fractional Calculus in Continuum Mechanics\/},
  Springer Verlag, Wien and New-York, 1997, pp. 291--248.
 [Reprinted in {\tt http://www.fracalmo.org}]


\bibitem{MaiGor JCAM00}   
  F. Mainardi and R. Gorenflo,
  On Mittag-Leffler type functions in fractional evolution processes,
  {\it J. Comput.   Appl. Math.} {\bf 118}  (2000) 283-299.


\bibitem{Mainardi LUMAPA01}
 F. Mainardi, Yu. Luchko and G. Pagnini,
     The fundamental solution of the space-time fractional diffusion
     equation,
   {\it Fract. Calc. Appl. Anal.}
 {\bf 4} (2001) 153-192.
 [E-print      {\tt http://arxiv.org/abs/cond-mat/0702419}]
\bibitem{MainardiPagnini AMC03}
F. Mainardi and G. Pagnini,
The Wright functions as solutions of the time-fractional diffusion equations,
{\it   Appl. Math.  Comput.} {\bf 141} (2003) 51-62.
\bibitem{Mainardi TEMME06}
 F. Mainardi and  G. Pagnini,
 The role of the Fox-Wright functions in fractional subdiffusion
of distributed  order,
 {\it   J. Comput.  Appl. Math.} {\bf 207} (2007) 245-257.
 [E-print      {\tt http://arxiv.org/abs/0711.3779}]



\bibitem{Marichev 83}
 O.I. Marichev,
 {\it Handbook of Integral Transforms of Higher Transcendental Functions,
 Theory and Algorithmic Tables},
 Chichester, Ellis Horwood, 1983.
\bibitem{Metzler PhysA94}   
 R. Metzler, W.G.  Gl\"ockle   and  T.F. Nonnenmacher,
   Fractional model equation for anomalous diffusion,
   {\em Physica A\/} {\bf 211} (1994)  13--24.

 \bibitem{Metzler-Klafter JPhysics04}
 R. Metzler and J. Klafter,
The restaurant at the end of the random walk: Recent developments
 in the description of anomalous transport by fractional dynamics,
 {\it J. Phys. A. Math. Gen.}  {\bf 37} (2004)  R161-R208.
\bibitem{Naber 04}
M. Naber,
 {Distributed order fractional subdiffusion},
  {\it Fractals} {\bf 12} (2004) 23-32.
\bibitem{Saichev PhysA05}
A. Piryatinska, A.I. Saichev and W.A. Woyczynski,
 Models of anomalous diffusion: the subdiffusive case,
 {\it Physica A}  {\bf 349} (2005)  375-420.
\bibitem{Podlubny 99}
  I. Podlubny,
  {\it Fractional Differential Equations},
  Academic Press, San Diego, 1999.
\bibitem{Saichev-Zaslavsky 97}
{A. Saichev and G. Zaslavsky},
Fractional kinetic equations: solutions and applications,
 {\em Chaos\/} {\bf 7} (1997)  753-764.
 \bibitem{SKM 93}
 S.G.  Samko, A.A. Kilbas and O.I. Marichev,
 {\it Fractional Integrals and Derivatives: Theory  and  Applications},
 Gordon and Breach, New York, 1993.
 Translation from the Russian edition,
 Nauka i Tekhnika, Minsk, 1987.
  \bibitem{SchneiderWyss 89}
  {W.R. Schneider and  W. Wyss},
  Fractional diffusion and wave equations.
  {\it J. Math. Phys.} {\bf 30} (1989) 134-144.
\bibitem{SokolovChechkinKlafter 04}  
 I.M. Sokolov, A.V. Chechkin and J. Klafter,
Distributed-order fractional kinetics,
{\it Acta Phys. Polon.} {\bf 35} (2004) 1323-1341.
\bibitem{SokolovKlafter EINSTEIN05}
 I.M. Sokolov and J. Klafter,
From diffusion to anomalous diffusion: a century after Einstein's
Brownian motion,
{\it Chaos} {\bf 15} (2005) 026103-026109.

 \bibitem{Srivastava-Saxena AMC03}
 H.M. Srivastava and R.K. Saxena,
 {Operators of fractional integration and their applications},
 {\it Appl. Math.  Comput.} {\bf 118} (2003) 1-52.


 \bibitem{Umarov-Gorenflo ZAA05}
 S. Umarov and R. Gorenflo,
 Cauchy and nonlocal multi-point problems for distributed order
pseudo-differential equations: Part one,
{\it Z. Anal. Anwendungen} {\bf 24} (2005) 449-466.
 \bibitem{Zaslavsky PhysRep02}
 G.M. Zaslavsky,
 Chaos,  fractional kinetics  and anomalous transport,
{\it Phys. Rep.}  {\bf 371} (2002)  461-580,


\end{thebibliography}
\end{document}